\documentclass[10pt]{iopart}
\usepackage{graphicx}

\begin{document}
\title[Atomic excitation during multi-electron tunnel
ionization]{On the recollision-free excitation of krypton during
ultrafast multi-electron tunnel ionization}

\author{W A Bryan$^{1}$, S L Stebbings$^{1\dag}$, J McKenna$^{2}$,
\\E M L English$^{1}$, M Suresh$^{2\ddag}$, J Wood$^{1}$, B Srigengan$^{2}$,
\\I C E Turcu$^{3}$, I D Williams$^{2}$ and W R Newell$^{1}$}

\address{1) Department of Physics and Astronomy, University College
London, Gower Street, London WC1E 6BT, UK}
\address{2) Department of Pure and Applied Physics, Queen's University Belfast,
Belfast BT7 1NN, UK}
\address{3) Central Laser Facility, CCLRC Rutherford Appleton Laboratory,
Chilton, Didcot, Oxon. OX11 0QX, UK}
\address{\dag Present address: Department of Physics and Astronomy,
University of Southampton, Southampton, SO17 1BJ, UK}
\address{\ddag Present address: Cavendish Laboratory, University of
Cambridge, Madingley Road, Cambridge CB3 0HE, UK}

\ead{w.bryan@ucl.ac.uk}

\begin{abstract}
The probability of multiple ionization of krypton by 50
femtosecond circularly polarized laser pulses, independent of the
optical focal geometry, has been obtained for the first time. The
excellent agreement over the intensity range 10 TWcm$^{-2}$ to 10
PWcm$^{-2}$ with the recent predictions of A. S. Kornev
$\it{et~al}$ [Phys. Rev. A {\bf 68}, 043414 (2003)] provides the
first experimental confirmation that non-recollisional electronic
excitation can occur in strong field ionization. This is
particularly true for higher stages of ionization, when the laser
intensity exceeds 1 PWcm$^{-2}$ as the energetic departure of the
ionized electron(s) diabatically distorts the wavefunctions of the
bound electrons. By scaling the probability of ionization by the
focal volume, we discusses why this mechanism was not apparent in
previous studies.
\end{abstract}

\section{Introduction}
Modern intense ultrafast pulsed lasers generate an electric field
of sufficient strength to tunnel ionize one or more valence
electrons from an atom. Such processes are generally treated as a
rapid succession of isolated events, in which the states of the
remaining bound electrons are unrelated to the preceding
ionization events. While such a description has been shown to be
more than adequate at predicting single, and in certain cases
double ionization, a number of recent experimental studies have
indicated the necessity to consider more than one active electron
\cite{yama}. In the current study, we present evidence for the
existence of indirect non-recollisional excitation of the bound
valence electrons during tunnel ionization (TI).

The foundation of the modern theoretical description of intense
AC-field nonlinear photoionization was laid by Keldysh
\cite{keld}, who derived the dependence of the rate of TI on the
frequency and strength of the optical field and the binding energy
and quantum state of the ion and electron: for a detailed review
see \cite{popov}. In the ultrafast regime (optical pulse duration
of the order of femtoseconds, 1 femtosecond = 10$^{-15}$ s), the
ionization of an atom is either a perturbative process (peak
intensity less that approximately 10 TWcm$^{-2}$ where 1 TW =
10$^{12}$ W), or a strong-field process, described by tunnel
theory. In the present work, we concern ourselves purely with the
strong field regime.

Immediately following ionization, the liberated electron is in a
modified Volkov state \cite{volk}, and is initially fully
correlated with the parent ion. The electron is driven on a
trajectory defined by the ellipticity of the laser radiation and
the optical phase at which ionization occurred. An intriguing
phenomenon in the strong field regime is that of
\emph{recollision} \cite{cork} whereby electron impact excitation
\cite{feuer} or further ionization \cite{mos} can arise in a
linearly polarized laser pulse, which drives electron(s) back to
the parent ion. The COLTRIMS technique \cite{colt} has recently
been employed to great effect to investigate these mechanisms.
Recollision is the key mechanism for coherent attosecond XUV pulse
generation \cite{reid}, whereby energy absorbed by the electron
from the field is dissipated photonically upon recombination with
the parent ion.

Tunnel theory, often the popular ADK form \cite{adk}, is unable to
quantify recollision, however by treating recollision ionization
(RI) as TI from the previous charge state, albeit with a
suppressed efficiency, an adequately accurate numerical solution
may be found \cite{laro,eich}. RI can be treated in a more
physically robust manner either through intense-field many-body
\textit{S}-matrix theory \cite{beck} or time-dependent density
functional theory \cite{tddft}. The time-dependent solution of the
Schrodinger equation \cite{tdse} provides the most accurate
description of the process at the expense of only being applicable
to two-electron systems and then with a massive computational
overhead.

In the present work however, we have made RI events extremely
unlikely by employing circularly polarized light. In undergoing
tunnel ionization, an atom must absorb a large number of photons
\cite{keld} which, when absorbed from a circularly polarized laser
pulse, transfer considerable angular momentum to the liberated
electron, dramatically reducing the probability of returning to
the ionic core, i.e. the impact parameter will be very large. In
terms of the laser-induced electric field, following TI, the field
drives the free electron on a spiral path: the pitch of the spiral
is defined by the temporal envelope of the pulse. As the pulse
intensity increases, the electron is rapidly removed from the
vicinity of the ion. By negating recollision processes, the
masking effects of electron-impact excitation and ionization are
removed, revealing the existence of a new laser-induced excitation
mechanism.

While ionization in ultrafast laser pulses is well documented,
minimal theoretical and essentially no experimental studies have
investigated the possibility of simultaneous excitation of the
parent ion during TI. The contemporary work of Zon \cite{zon}
introduced the idea of `inelastic tunnelling' whereby the parent
ion can be left in an excited state following the ionization of
one of $\it{N}$ identical valence electrons. The excitation
process is through `shake-up', first employed by Carlson
\cite{carl} to explain single UV photon absorption leading to the
ionization of a first electron with the excitation of a second
electron. The ionization event diabatically distorts the bound
electron wavefunctions, resulting in the excitation of a bound
electron.

Zon \cite{zon} and Kornev \textit{et al} \cite{korn} have derived
a general expression for the rate of TI of an atom with
simultaneous excitation of the lowest lying ionic states. This
derivation relies on the sudden approximation, valid in the case
of ultrafast TI as the ionization potential is far greater than
the excitation energy of the ion. The range of excited states of
Kr$^{m+}$ ($\it{m}$ = 2 to 6) considered is consistent with this
approximation, modifying the probability of ionization
significantly. Furthermore, it is also necessary to allow for all
combinations of excitation + ionization to some final charge state
after the laser pulse has finished, irrespective of whether it is
a ground or excited state. To distinguish such processes from
standard sequential TI, we refer to such processes as
multi-electron tunnel ionization (METI). This distinction is
necessary, as Eichmann \textit{et al} \cite{eich} have previously
proposed a mechanism of collective tunnel ionization (CTI),
whereby two electrons can simultaneously tunnel away from the
atom. This process was predicted to only arise if the tunnelling
electrons have highly correlated momenta, otherwise recapture of
one of the electrons would occur; this requirement results in a
very low probability.

\section{Experimental Procedure}
In general, previous experimental measurements of ion yield as a
function of laser intensity are unavoidably a convolution of the
probability of ionization with the focal volume producing the
signal \cite{peg,hansch}. By simply changing the energy of the
laser pulse, the spatial distribution of laser intensity also
changes, resulting in the characteristic
$\it{I}^{3\hspace{-2pt}/\hspace{-2pt}2}$ response at intensities
above saturation, the so-called `volume variation' problem. While
most apparent above saturation, there is also an intrinsic volume
dependence below saturation, which subtly modifies the gradient of
the ion yield. Frequently, the complexity of this situation is
compounded by diffraction associated with the spatial profile of
the laser; a direct comparison with theory being made impossible
without introducing the specific experimental geometry. Through a
contemporary method by Bryan \textit{et al} \cite{bryan},
developed from the pioneering work of Walker \textit{et al}
\cite{walk} and analogous to the tomographic technique of
Goodworth \textit{et al} \cite{good}, we circumvent this
hinderance.

A novel solution to the volume variation problem was proposed by
Van Woerkom and co-workers \cite{hansch}: by softly focusing a
high-energy ultrafast laser pulse into a tightly apertured
photoion detector, only those ionization states generated within a
narrow spatial (and therefore intensity) window are detected. By
translating the focusing optic, this restricted range of observed
laser intensities may be accurately manipulated.

Up to six-fold ionization of krypton saturates at intensities less
than 10 PWcm$^{-2}$, and the 30 mJ 790 nm 50 fs laser pulses
generated by the ASTRA Laser Facility (UK) need only be softly
focussed (\textit{f}/11 optics) to generate a peak intensity in
excess of 100 PWcm$^{-2}$. Indeed the active range of the focused
ASTRA beam extends over tens of millimetres. By limiting the
spatial acceptance of our ion time-of-flight mass spectrometer to
250 microns, the ion yield is strongly dependent on the position
of the optical focus with respect to the spectrometer. By
recording the relative ion yield of Kr$^{n+}$ ($\it{n}$ = 1 to 6)
while translating the focusing optic in 125 micron steps, the
intensity selective scan presented in figure 1 is measured
\cite{bryan,walk,good}. Throughout this measurement, the krypton
gas pressure is low enough so as to avoid space-charge effects,
tested by repeating the intensity selective scan (ISS)
measurements as a function of pressure. At higher number
densities, space-charge effects are apparent as a loss of ion
yield definition both in time-of-flight and focal position.

\begin{figure}
\begin{center}
\includegraphics[width=300pt]{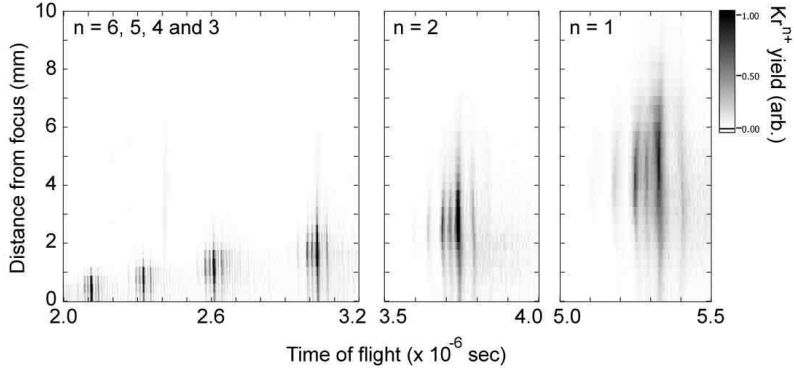}\\
\caption{Intensity selective scan ion yield matrix for Kr$^{n+}$
(\textit{n} = 1 to 6) generated at the focus of a 50fs 790nm
circularly polarized laser pulse. The 2000-laser shot average ion
yield is recorded with an apertured time-of-flight mass
spectrometer as the focusing optic is translated parallel to the
direction of propagation. The six most abundant isotopes are
clearly resolved.}\label{fig1}
\end{center}
\end{figure}

At this point, we wish to stress the importance of identifying
background contaminants in the ISS data. In previous studies on
argon \cite{bryan2}, such an incumbrance is not present, as argon
has three naturally occurring isotopes $^{36}$Ar, $^{38}$Ar and
$^{40}$Ar with abundances of 0.33\%, 0.06\% and 99.60\%
respectively, which do not suffer from significant charge-to-mass
degeneracy with any possible atmospheric contaminants (greater
than 0.1\% yield). Krypton however has six main isotopes
$^{78}$Kr, $^{80}$Kr, $^{82}$Kr, $^{83}$Kr, $^{84}$Kr and
$^{86}$Kr, with natural abundances 0.35\%, 2.28\%, 11.58\%,
11.49\%, 57.00\% and 17.30\% respectively. The high charge-to-mass
ratio resolution of our spectrometer allows selective ion yield
integration over those isotopes not degenerate with background
contaminants. An example is the Kr$^{3+}$ peak in figure 1:
$^{84}$Kr$^{3+}$ is degenerate with N$_2^+$. Ionization to N$_2^+$
requires a far lower intensity than $^{84}$Kr$^{3+}$, apparent in
figure 1 as the faint ion yield extending to large distances from
the focus. To recover the probability of ionization, the
non-degenerate isotopes (78, 80, 82 and 86) are integrated with
respect to flight time, thus the spatially dependence of the ion
yield is measured.

To make the ISS measurements directly comparable with theoretical
predictions, we remove the spatial integration through a
deconvolution \cite{bryan,walk} or inversion \cite{good}
technique, requiring the measured ion yield and theoretical
on-axis intensity as a function of focal position as inputs. The
resulting partial probability of ionization (PPI) \cite{bryan} is
independent of the effects of varying signal-producing volume. The
unavoidable diffraction of the laser pulse is also accounted for
by solving the Collins form \cite{collins} of the Huygens-Fresnel
diffraction integral \cite{huygens}. The enhancement of detector
gain with charge state (quantum efficiency) is also quantified.
The PPI is therefore equivalent to the response of a single atom
to a spatially infinite laser focus, as we have removed \emph{all}
instrumental dependence. The PPI results are valid up to
saturation, however at higher intensities the deconvolution breaks
down, thus the PPI is then defined as unity.

\begin{figure}
\begin{center}
\includegraphics[width=180pt]{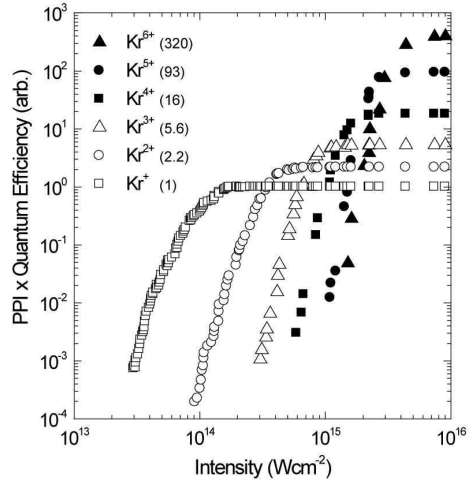}\\
\caption{Partial probability of ionization (PPI) to Kr$^{n+}$
(\textit{n} = 1 to 6) as a function of laser intensity recovered
from the ISS data presented in figure 1. The deconvolution of a
non-Gaussian focal volume as detailed in Bryan \textit{et al}
\cite{bryan} results in a PPI which is independent of the spatial
distribution of laser intensity across the laser focus. The PPI at
which saturation occurs relative to the PPI of Kr$^+$ is
proportional to the quantum efficiency of the detector. The values
in parenthesis in the legend indicate the relative efficiency for
each charge state.}\label{fig2}
\end{center}
\end{figure}

The raw PPI(\textit{n}) for Kr$^{n+}$ (\textit{n} = 1 to 6)
results are presented in figure 2. The influence of the variation
of quantum efficiency of our detector is apparent from the
increasing PPI(\textit{n}) at which saturation occurs with
increasing ion charge state. In figure 2, the values in
parenthesis in the legend are the quantum efficiencies relative to
the PPI of Kr$^+$. If the PPI(\textit{n}) for Kr$^{n+}$
(\textit{n} = 1 to 6) presented in figure 2 is divided by the
quantum efficiency, all PPI(\textit{n}) saturate to unity at high
intensity.

\begin{figure}
\begin{center}
\includegraphics[width=320pt]{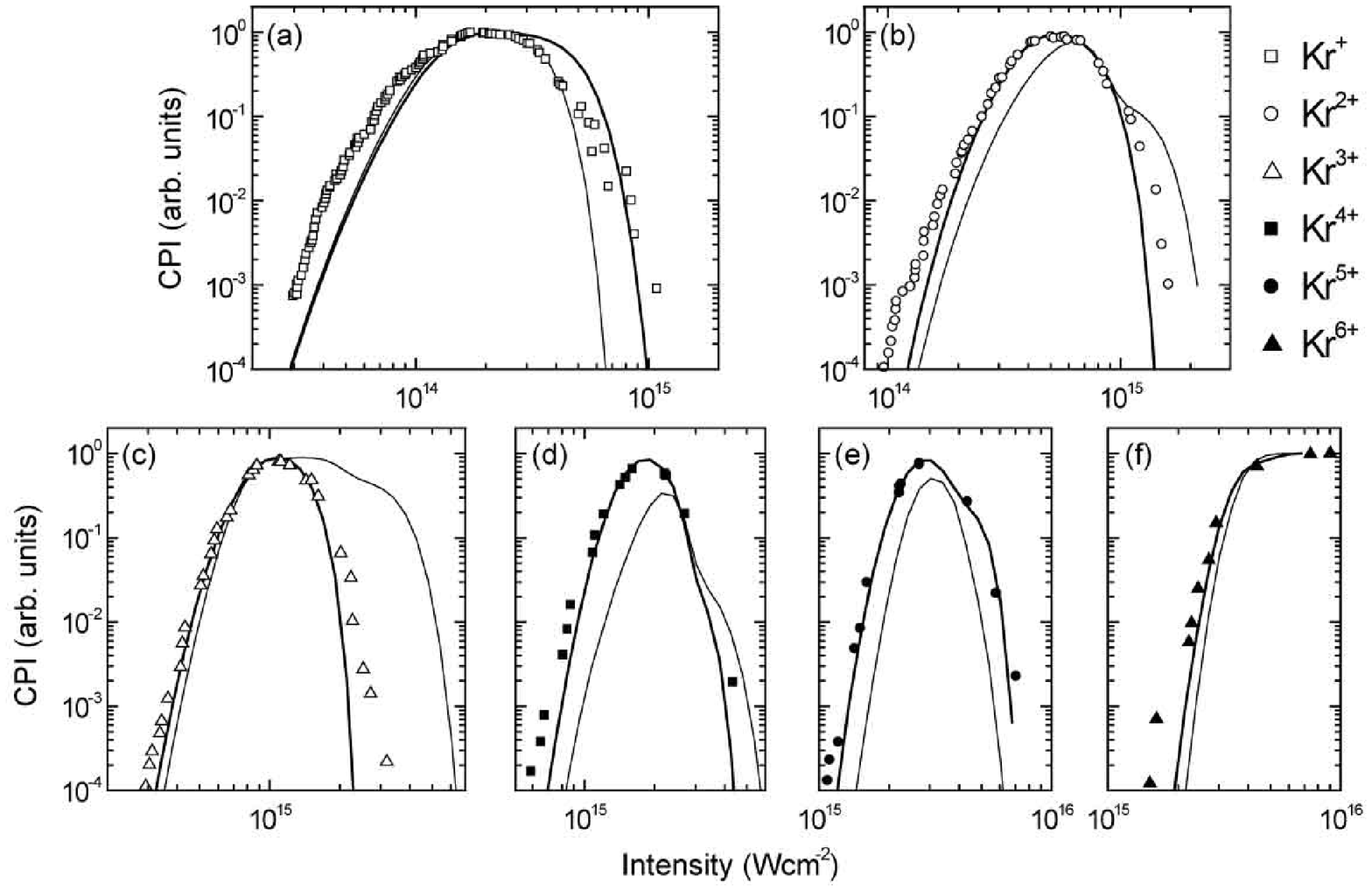}\\
\caption{Conserved probability of ionization (CPI) to Kr$^{n+}$
(\textit{n} = 1 to 6) by a 50fs 790nm circularly polarized laser
pulse. All influence of both the optical or detector geometry has
been removed, allowing a direct comparison with the predictions of
Kornev \textit{et al} \cite{korn}. In all frames (a) to (f), the
thin line is traditional sequential TI (ADK \cite{adk})
prediction, while the thick line is the CPI for multi-electron
tunnel ionization (METI) including an allowance for excitation to
low lying states. The remarkable agreement illustrates the
breakdown of tunnel theory, particularly for triple (c) or higher
order ionization, where the agreement between the measured CPI and
the METI prediction indicates the presence of
excitation.}\label{fig3}
\end{center}
\end{figure}

If the quantum efficiency of the detector is removed, such that
the PPI(\textit{n}) are now normalized, the break down of the
deconvolution routine at intensities greater than saturation can
be addressed by conserving the probability of ionization. Consider
the PPI(\textit{n}) for \textit{n} = 1, 2 as shown in figure 2. At
low intensity, PPI(1) is small, and increases with intensity up to
saturation at an intensity of 200 TWcm$^{-2}$. However, as is
often the case in atomic ionization, PPI(2) is nonzero at this
intensity. The condition for conserving probability is that the
sum of probabilities is less than unity below the saturation of
PPI(1) or equal to unity at higher intensities, thus at an
intensity greater than 200 TWcm$^{-2}$, the conserved probability
of ionization to Kr$^+$, CPI(1) must be less than unity as PPI(2)
is nonzero. This definition is extended to an \textit{N} electron
system in the present work and is only valid following the removal
of the quantum efficiency.

\section{Results}
A direct comparison between the theoretical predictions of Kornev
\textit{et al} \cite{korn} and the present volume-independent CPI
is presented in figure 3. All theoretical curves presented are
universally normalized in intensity to the experimental data. In
all frames of figure 3, the sequential ADK prediction is the thin
line, and the multi-electron tunnel ionization (METI) predication
is the thick line. In figure 3(a), neither the ADK or METI
predictions represent the Kr$^+$ CPI: this is due to a
contribution from multiphoton ionization (MPI) at sufficiently low
intensities to access the perturbative regime. At intensities
close to the saturation of Kr$^+$, it can be seen that both ADK
and METI describe the CPI adequately. A similar low intensity
response is observed in figure 3(b) (Kr$^{2+}$), where again
around 100 TWcm$^{-2}$ MPI contributes. However, as the intensity
increases, the METI prediction is in much better accord with the
experimental data, reproducing the CPI far more closely than ADK.
The superiority of agreement with METI over ADK is dramatically
illustrated for ionization to Kr$^{3+}$, Kr$^{4+}$ and Kr$^{5+}$,
figure 3(c) to (e), particularly high intensity Kr$^{3+}$ and
Kr$^{4+}$ at all intensities. Indeed, at an intensity of around 1
PWcm$^{-2}$, ADK theory underestimates the CPI of Kr$^{4+}$ by
more than an order of magnitude. At the highest intensities
discussed in the present work, ionization to Kr$^{6+}$, figure
3(f) is well described by both ADK and METI, however the latter is
still observed to give the better agreement. Throughout figure 3,
it is apparent that METI generates an excellent quantification of
the CPI of krypton by circularly polarized 50 fs laser pulses.
This is a direct consequence of the excitation of the bound
valence electrons in the atomic ion. The shake-up mechanism
invoked in both the present work and \cite{korn} is the subject of
ongoing interest, see for example the recent review by Becker
\textit{et al} \cite{beck}: shaking can only cause excitation at
high intensities, as the energy of the departing electron must be
sufficient to excite other residual electrons. This is supported
by the present work: there is no significant excitation until
around 1 PWcm$^{-2}$. Becker \textit{et al} \cite{beck} predict
many orders of magnitude difference between the recollision yield
and shake up contribution. However, here the recollision
contribution is negligible. In qualitative terms the high degree
of agreement between the measured CPI and the METI prediction is a
result of (i) the high intensity laser-induced population of
low-lying excited states through diabatic shake-up excitation,
(ii) the removal of recollision excitation or ionization through
the use of circular polarization. Furthermore, this observation
would not be possible without the measurement of the
geometry-independent PPI and the realization of conservation of
probability. The pertinent question raised by the present work is
why has this mechanism not been observed before?

\begin{figure}
\begin{center}
\includegraphics[width=260pt]{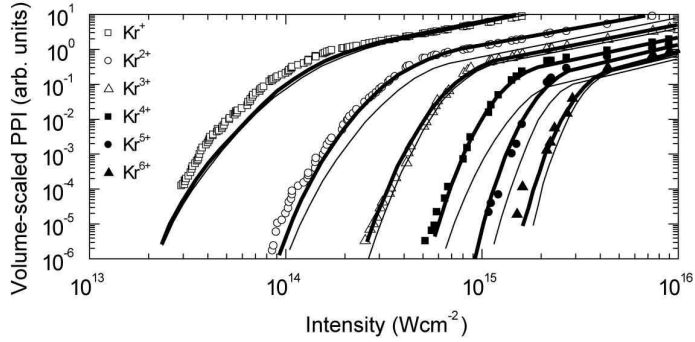}\\
\caption{Volume-scaled partial probability of ionization to
Kr$^{n+}$ for \textit{n} = 1 to 6 as compared to the volume-scaled
predictions of Kornev \textit{et al} \cite{korn}, illustrating how
ionization-induced excitation could be more difficult to identify
from traditional intensity variation measurements. As in figure 3,
the thin line is traditional sequential ADK prediction, while the
thick line is multi-electron tunnel ionization (METI) prediction,
allowing the generation of excited states by
ionization.}\label{fig4}
\end{center}
\end{figure}

\section{The presence of excitation in previous experiments}
In figure 4, we present the PPI(\textit{n}) to Kr$^{n+}$ for
\textit{n} = 1 to 6, however in contrast to figure 2, the PPI for
each ionization state is scaled by the volume of the laser focus
generating the signal. This is easily quantified, as the
deconvolution route requires the precise computation of the
spatial distribution of intensity, as described in \cite{bryan}.
Figure 4 also contains the volume-scaled METI predictions
\cite{korn}. As is expected, the observations of the previous
section are still applicable, however it is now far more difficult
to isolate which model applies above saturation.

Comparing figures 3 and 4, we illustrate how subtle differences
between the volume-scaled PPI and either theoretical prediction
could be missed in a traditional intensity-variation measurement
(equivalent to figure 4, for example see \cite{laro}). If the
detector efficiency is unknown, the measurement of which is by no
means trivial, the effect of excitation could conceivably be
overlooked as a manifestation of detector efficiency or could be
ignored if the ADK curves are normalized independently. As
apparent from figure 4, excitation is manifest as a major vertical
shift in apparent yield, and a minor variation in intensity
response. Furthermore, a number of experimental studies have
commented on the inadequacy of tunnel theory (specifically the ADK
treatment \cite{adk}), without concrete discussion of why, nor the
suggestion of a mechanism by which tunnel theory might be breaking
down. It is hoped that the present work allows previous
experimental data to be re-examined in terms of the
ionization-induced excitation mechanism.

\section{Conclusion}
We present for the first time strong evidence for the presence of
considerable atomic excitation during tunnel ionization of krypton
by a 790nm 50fs circularly polarized laser pulse focused to
intensities in excess of 10 TWcm$^{-2}$. The polarization of the
radiation is such that recollision excitation and ionization are
essentially negated. The impressive agreement between the measured
conserved probability of ionization and recent theoretical
predictions indicate that excitation during ionization need be
considered irrespective of recollision processes. Excitation is
due to the intense laser field energetically removing valence
electrons: during these tunnel ionization events, the
wavefunctions of the remaining electrons is impulsively distorted.
Such excitation is also expected to occur in a linearly polarized
laser field, and, as the results of Kornev \textit{et al}
\cite{korn} suggest, is expected to be even more important in a
5fs laser pulse. This has a major bearing on the emerging field of
optical attosecond physics. The influence of initial and
transitionary electronic states accessed through intra-pulse
excitation must be quantified to accurately predict the energy of
the emitted photons.

\ack{This work is supported by the EPSRC, UK. Research
studentships are acknowledged by SLS, EMLE and JW (EPSRC), JMcK
(DEL) and MS (IRCEP at QUB). The authors would like to thank the
following staff of the ASTRA Laser Facility (Rutherford Appleton
Laboratory, UK): J M Smith, E J Divall, C J Hooker and A J
Langley. The authors would also like to acknowledge A S Kornev and
B A Zon from Voronezh State University, Russia for electronically
communicating their theoretical data and for highly fruitful
discussions.}

\end{document}